\documentclass[twocolumn, prb, aps, showpacs, letterpaper]{revtex4}

\usepackage[dvips]{graphicx} 
\usepackage{subfigure}
\usepackage{latexsym}
\usepackage{dcolumn} 
\usepackage{bm} 

\begin{document}
\title{Correlations between pressure and bandwidth effects in metal-insulator transitions in manganites}

\author{Congwu Cui}
\author{Trevor A. Tyson}
\affiliation{Physics Department, New Jersey Institute of Technology, Newark, New Jersey 07102}

\date{\today}

\begin{abstract}
The effect of pressure on the metal-insulator transition in
manganites with a broad range of bandwidths is investigated. A critical
pressure is found at which the metal-insulator transition temperature, T$_{MI}$, reaches a maximum value in every sample studied. The origin of this universal pressure and the relation between the pressure effect and the bandwidth on the metal-insulator transition are discussed.
\end{abstract}



\maketitle


Manganites have been the focus of intense studies in recent years since the observation of large magnetoresistance sparked interest in these materials for use as magnetoresistance sensors.  The metal-insulator transitions (MIT) observed in this kind of materials is crucial to the colossal magnetoresistance effect. Metal-insulator transition occurs in manganites in two cases: first, metallic ground  state exists in the low temperature range in some doping systems at certain doping concentrations, such as: La$_{1-x}$Sr$_{x}$MnO$_{3}$ (x$\sim$0.16-0.50), La$_{1-x}$Ca$_{x}$MnO$_{3}$ (x$\sim$0.18-0.50), Nd$_{1-x}$Sr$_{x}$MnO$_{3}$ (x$\sim$0.25-0.50); second,  metallic states can be induced by other factors, such as magnetic fields, photons, pressure, electric fields.  Pr$_{1-x}$Ca$_{x}$MnO$_{3}$ at x$\sim$0.3 is typical in the latter class. In most of the manganites, the metallic state is coupled to the ferromagnetic state so that the very large magnetoresistance can be explained by double exchange theory.

In the parameters determining the complicated properties of the manganites, the e$_{g}$ electron bandwidth W is a particularly important one to the metal-transition temperature T$_{MI}$, or the appearance of the metallic state under some factors. In manganites, the Mn 3d orbital is split into t$_{2g}$ and e$_{g}$ orbitals by the octahedral crystal field. The conduction band electrons are of e$_{g}$ symmetry. Because the e$_{g}$ orbital is Jahn-Teller active, Jahn-Teller distortion (JTD) can further split the two-fold degenerate e$_{g}$ orbital to trap the  conduction band electrons. Consequently, the bandwidth is highly correlated with the local atomic structure of the MnO$_{6}$ octahedra: cooperative tilting (Mn-O-Mn bond angle), Jahn-Teller distortion (Mn-O distances) and coherence of the JTD. The bandwidth is characterized by the overlap between the Mn-3d orbital and O-2p orbital and can be described empirically by an equation:\cite{medarde_prb_52_9248_95}
\begin{equation}\label{equ_1}
    W \propto \frac{{\cos [\frac{1}{2}(\pi  -  < \beta  > )]}}{{d_{Mn-O}^{3.5} }}
\end{equation}
where W is the bandwidth, $\beta$ is the Mn-O-Mn bond angle, and d$_{Mn-O}$ is the Mn-O bond length.  In double exchange theory, it is described as the electron hopping rate or the transfer integral: $t_{ij}  = t_{ij}^0 \cos (\frac{{\theta _{ij} }}{2})$, where  $t_{ij}^{0}$ is the transfer integral that depends on the spatial wave function overlaps, $\theta_{ij}$ is the relative angle between two neighboring Mn ion t$_{2g}$ core spins.

Generally, the structure can be tuned in two ways: chemical doping and external pressure. In chemical doping, by selecting different doping elements and doping concentration, the average A-site atom size $<$r$_{A}$$>$ in the AMn$O_{3}$ system is changed.  Because of the mismatch between $<$r$_{A}$$>$ and Mn-site ion size, the local atomic structure of MnO$_{6}$ octahedra can be modified. Therefore, the bandwidth is tuned by chemical doping so that complicated electronic and magnetic phase diagrams have been observed.\cite{tokura_jmmm_200_1_99} The external pressure method, is a ``clean method'' that only modifies lattice structure without inducing chemical complexity. To date, in studies on manganites, the effects of external pressure on the charge ordering, metal-insulator transition, magnetic states have been observed.

Currently, most of the high pressure studies on manganites are on metal-insulator transition. In the low pressure range, this electronic transition is coupled to the ferromagnetic transition, which can be explained qualitatively by double exchange theory. \cite{neumeier_prb_52_R7006_95, moritomo_prb_55_7549_97} It is also found that hydrostatic pressure has a similar effects to chemical doping with larger atoms and higher doping concentration. Both can increase the Mn-O-Mn bond angle, compress Mn-O bond length and hence, lead to larger bandwidth. Correspondingly, T$_{C}$ (or T$_{MI}$) increases, or in some manganites originally in insulating state, a MIT is induced. The effect of chemical doping and pressure can be scaled to each other with a conversion factor 3.75$\times$10$^{-4}$ \AA/kbar.\cite{hwang_prb_52_15046_95}  However, most pressure experiments were conducted below 2 GPa.

By applying pressures up to $\sim$6 GPa on  manganite systems with a broad range of bandwidths, the effect of pressure on the MIT and the correlation between the pressure effect and the chemical doping were observed. It is found that T$_{C}$ and/or T$_{MI}$ do not change monotonically with pressure \cite{cui_prb_67_104107_03} and these two transitions do not always couple.\cite{cui_apl_83_2856_03} A universal pressure may exist for the metal-insulator transition in manganites. With an increase of the bandwidth, the change in the metal-insulator transition temperature with pressure may vanish.


To systemically explore the external pressure effect and the chemical doping effect and the correlation between them, several manganite systems with much different ground magnetic and electronic states, Nd$_{1-x}$Sr$_{x}$MnO$_{3}$ (x = 0.45, 0.50) (NSMO), La$_{0.60}$Y$_{0.07}$Ca$_{0.33}$MnO$_{3}$ (LYCMO), and Pr$_{1-x}$Ca$_{x}$MnO$_{3}$ (x = 0.25, 0.30, 0.35) (PCMO), were selected. The bandwidth of the selected samples uniformly distributes in the electronic phase diagram  (see Ref.~\onlinecite{hwang_prb_52_15046_95}).

The samples were prepared by solid-state reaction. The procedure and details of making the samples were described elsewhere.\cite{cui_prb_67_104107_03,cui_manuscript_pcmo1,cui_manuscript_nsmo} All the samples are characterized with the x-ray diffraction and magnetization measurements. The details of high pressure resistivity measurement method and error analysis were described previously.\cite{cui_prb_67_104107_03} The metal-insulator transition temperature whenever present is defined as the temperature at the resistivity peak. Because of the lower temperature stability of our system in the cooling cycle, the data were taken only while warming up.


In all the samples studied, there is a MIT at ambient pressure or a MIT can be induced by applying pressure. Corresponding to the bandwidth phase diagram in Ref.\ \onlinecite{hwang_prb_52_15046_95}, the Nd$_{1-x}$Sr$_{x}$MnO$_{3}$ (x = 0.45, 0.50) system has a large bandwidth; La$_{0.60}$Y$_{0.07}$Ca$_{0.33}$MnO$_{3}$ has a medium bandwidth; Pr$_{1-x}$Ca$_{x}$MnO$_{3}$ (x = 0.25, 0.30, 0.35) system has a small bandwidth.

\begin{figure}
\includegraphics[width=3in]{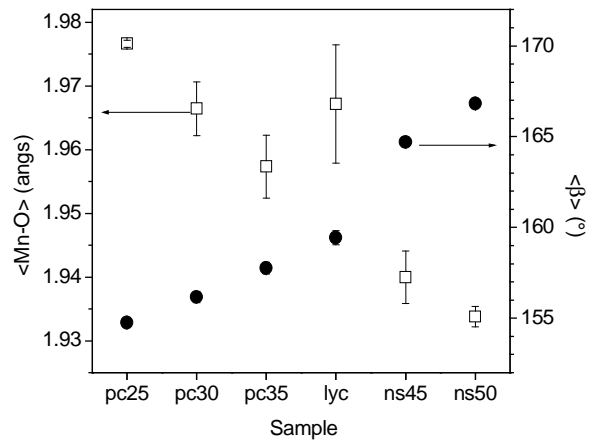}\\
\caption{\label{fig1}Mn-O bond length and Mn-O-Mn bond angle of samples at ambient conditions. Note: the abscissa gives the samples: the numbers indicated the concentration of the doping element; p represents Pr, c represents Ca, l represents La, y represents Y, n represents Nd, s represents Sr.}
\end{figure}

In Table \ref{table1}, $<$r$_{A}$$>$, the tolerance factor t, and metal-insulator transition temperature at ambient pressure, which corresponds to the bandwidth, are listed. The average Mn-O bond length and Mn-O-Mn bond angle of all samples determined from the Rietveld refinement to the x-ray diffraction patterns are shown in Fig.~\ref{fig1}. According to equation (\ref{equ_1}), with increasing $<$r$_{A}$$>$ or t, the decreasing bond length and bond angle lead to increasing bandwidth W and hence, increasing T$_{MI}$.

\begingroup
\squeezetable
\begin{table}
\caption{\label{table1}Average A-site ion size, tolerance factor, T$_{MI}$, dT$_{MI}$/dP, and critical pressure}
\footnotetext{Note: t is the tolerance factor calculated with the data in Ref.~\onlinecite{shannon_aca_32_785_76}; T$_{MI}$ is the metal-insulator transition temperature at ambient pressure; dT$_{MI}$/dP is the change rate of T$_{MI}$ at P$\sim$0 extracted by fitting the data with a third-order polynomial, the numbers in brackets are the errors in the last one or two digits; P* is the pressure where the T$_{MI}$ increase trend reverses.}
\begin{ruledtabular} 
\begin{tabular}{lccccc}
  Sample&$<$r$_{A}$$>$&t&T$_{MI}$ (K)&dT$_{MI}$/dP (K/GPa)&P* (GPa)\\
  \hline
Pcmo25 & 1.17925 & 0.92711 & N/A & 51(7) & 3.8(9)\\
Pcmo30 & 1.17930 & 0.92830 & N/A & 88(7) & 3.5(4)\\
Pcmo35 & 1.17935 & 0.92950 & N/A & 68(7) & 3.9(3)\\
Lycmo  & 1.20230 & 0.93730 & 149(1) & 22(4) & 3.8(4)\\
Nsmo45 & 1.22915 & 0.94987 & 291(1) & 18(11) & 2.6(1.6)\footnote{The resistivity in paramagnetic phase at $\sim$316K gives a P* of 3.8$\pm$0.3 GPa.\cite{cui_manuscript_nsmo} See text for details.}\\
Nsmo50 & 1.23650 & 0.95374 & 256(1) & -1.0(2.8) & 3.0(1.6)\\
\end{tabular}
\end{ruledtabular}


\end{table}

\endgroup

\begin{figure}
\includegraphics[width=3in]{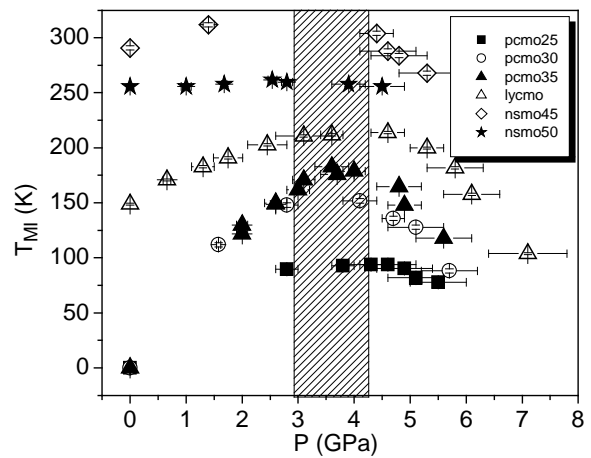}\\
\caption{\label{fig2}Pressure dependence of metal-insulator transition temperatures of Nd$_{1-x}$Sr$_{x}$MnO$_{3}$ (x = 0.45, 0.50); La$_{0.60}$Y$_{0.07}$Ca$_{0.33}$MnO$_{3}$; Pr$_{1-x}$Ca$_{x}$MnO$_{3}$ (x = 0.25, 0.30, 0.35).}
\end{figure}

With the application of  pressure, the metal-insulator transition temperatures of the samples which have MIT at ambient pressure increases. In the narrow bandwidth Pr$_{1-x}$Ca$_{x}$MnO$_{3}$ system, the samples are insulating at ambient pressure. Under pressure, metal-insulator transitions are induced. With pressure increase, the behavior of the T$_{MI}$ is similar to other samples with larger bandwidth. When the pressure is above a certain point, the increasing trend of T$_{MI}$ of all samples is reversed. The evolution of the transition temperatures of all samples are shown in Fig.\ \ref{fig2}.

In Fig.\ \ref{fig2}, the most salient feature is that a critical pressure P* exists in each sample: with pressure increase, below P*, T$_{MI}$ increases; above P*, T$_{MI}$ decreases. By fitting T$_{MI}$ \textit{vs}.\ P plots with a third-order polynomial, P* for each sample can be extracted and is listed in Table \ref{table1}. In the fitting error range, the samples have the same critical pressure P* (One exception is the NSMO45 sample, the P* determined from the T$_{MI}$ is small. But if we look at the resistivity changes with pressure, its critical pressure is the same as other samples. This may comes from the highest temperature limit of the instruments which leads that the T$_{MI}$ near to the limit in the middle range pressure can not be determined.\cite{cui_manuscript_nsmo}).

In large bandwidth samples, the change of T$_{MI}$ with pressure is slower than in narrow bandwidth ones, indicating that the large bandwidth samples are more stable under pressure. The samples studied are selected with different doping concentration and from different doping systems. The bandwidths spans a large range. The samples also have much different ground state electronic and magnetic properties at ambient conditions. But the metal-insulator transitions in these samples all follow a similar behavior, therefore, it is reasonable to speculate that the critical pressure P* is universal for the metal-insulator transitions in manganites. From the structural measurements on manganites,\cite{cui_prb_67_104107_03,congeduti_prl_86_1251_01, meneghini_prb_65_012111_02} the behavior of T$_{MI}$ under pressure could possibly be ascribed to a local atomic structure transformation of the MnO$_{6}$ octahedra.

Under pressure, the smaller bandwidth samples seem to have smaller pressure range in which the sample is metallic in low temperature and outside which they are insulating. On the other hand, the samples with large bandwidth is more stable under pressure and the variable range of T$_{MI}$ is small and they do not become insulating in a larger pressure range. The lower stability of T$_{MI}$ in small bandwidth samples may come from the small A-site atoms, which leave more space between the octahedra for them to rotate - accordingly, a smaller pressure window for the metallic state.

In Fig.\ \ref{fig2}, the large bandwidth samples have higher T$_{MI}$. The only exception is that in the Nd$_{1-x}$Sr$_{x}$MnO$_{3}$ system, the x = 0.5 compound nominally has a larger bandwidth than the x = 0.45 compound but has lower T$_{MI}$. This possibly results from the strong charge ordering effect in Nd$_{0.5}$Sr$_{0.5}$MnO$_{3}$.

The changing rate of the metal-insulator transition temperature with pressure at the ambient pressure, dT$_{MI}$/dP at P = 0, is also interesting. The values of dT$_{MI}$/dP extracted from the third-order polynomial fitting results are listed in Table~\ref{table1}. Clearly, the smaller the bandwidth, the larger dT$_{MI}$/dP, indicating that the local structure of smaller bandwidth sample is more distorted and has a relatively large degree to which it can be compressed by pressure.


In summary, by applying external pressure on manganites of different chemical doping system and doping concentration and hence different e$_{g}$ electron bandwidth, it is found that the pressure effect on the metal-insulator transition in manganites is not equivalent to that of the chemical doping. Only at low pressures, is the pressure effect on the metal-insulator transition analogy to the chemical doping with elements with large atom size. With pressure increase, the trend of T$_{MI}$ increase with pressure is reversed at a critical pressure, above which the transition temperature decreases with pressure and finally, the material may become insulating. The critical pressure is found to exist in all the samples studied and possibly is universal for the metal-insulator transition in the manganites. The bandwidth (chemical doping) determines how stable the material may be under pressure. The larger bandwidth manganites are more stable under pressure and therefore, have smaller dT$_{MI}$/dP near ambient pressure and smaller T$_{MI}$ variation under pressure. Because of the importance of the local atomic structure of the MnO$_{6}$ octahedra to the electronic and magnetic properties of the manganites, this work may also contribute to understanding the properties of thin films of manganites which are important in technological applications.


This work is supported by National Science Foundation Grant DMR-0209243.


\bibliography{bib_cuicw}
\end{document}